\documentclass[11pt]{article}
\usepackage{cite,amsmath,amssymb,epsf}

\def\nn{\nonumber}

\let\bm=\bibitem

\newcommand{\be}{\begin{equation}}
\newcommand{\ee}{\end{equation}}
\def\ba{\begin{array}}
\def\ea{\end{array}}
\def\ft#1#2{{\textstyle{\frac{\scriptstyle #1}{\scriptstyle #2}}}}
\def\fft#1#2{\frac{#1}{#2}}
\def\del{\partial}

\def\sst#1{{\scriptscriptstyle #1}}

\def\td{\tilde}
\def\wtd{\widetilde}
\def\ie{\rm i.e.\ }
\def\dalemb#1#2{{\vbox{\hrule height .#2pt
        \hbox{\vrule width.#2pt height#1pt \kern#1pt
                \vrule width.#2pt}
        \hrule height.#2pt}}}

\newcommand{\hoch}[1]{$\, ^{#1}$}
\newcommand{\bea}{\begin{eqnarray}}
\newcommand{\eea}{\end{eqnarray}}

\def\0{{\sst{(0)}}}
\def\1{{\sst{(1)}}}
\def\2{{\sst{(2)}}}
\def\3{{\sst{(3)}}}
\def\4{{\sst{(4)}}}
\def\5{{\sst{(5)}}}
\def\6{{\sst{(6)}}}
\def\7{{\sst{(7)}}}
\def\8{{\sst{(8)}}}

\def\im{{{\rm i}}}

\def\ep{{\epsilon}}

\def\CP{{{\mathbb C}{\mathbb P}}}

\def\gcd{{\rm gcd}}

\begin{document}
\begin{flushright}
MIFP-05-37 \\
{\bf hep-th/0512306}\\
December\  2005
\end{flushright}

\begin{center}

{\Large {\bf A New Construction of Einstein-Sasaki Metrics in $D\ge 7$}}

\vspace{20pt}

H. L\"u\hoch{\dagger}, C.N. Pope\hoch{\dagger 1} and J.F.
V\'azquez-Poritz\hoch{\ddagger 2}

\vspace{20pt}

{\hoch{\dagger}\it George P. \&  Cynthia W. Mitchell Institute
for Fundamental Physics,\\
Texas A\&M University, College Station, TX 77843-4242, USA}

\vspace{10pt}

{\hoch{\ddagger} \it Department of Physics,\\
University of Cincinnati, Cincinnati OH 45221-0011, USA}

\vspace{40pt}

\underline{ABSTRACT}
\end{center}

   We construct explicit Einstein-K\"ahler
metrics in all even dimensions $D=2n+4 \ge 6$, in terms of a
$2n$-dimensional Einstein-K\"ahler base metric. These are
cohomogeneity 2 metrics which have the new feature of including a
NUT-type parameter, in addition to mass and rotation parameters.
Using a canonical construction, these metrics all yield
Einstein-Sasaki metrics in dimensions $D=2n+5 \ge 7$. As is
commonly the case in this type of construction, for suitable
choices of the free parameters the Einstein-Sasaki metrics can
extend smoothly onto complete and non-singular manifolds, even
though the underlying Einstein-K\"ahler metric has conical
singularities.  We discuss some explicit examples in the case of
seven-dimensional Einstein-Sasaki spaces.  These new spaces can
provide supersymmetric backgrounds in M-theory, which play a
r\^ole in the AdS$_4$/CFT$_3$ correspondence.

{\vfill\leftline{}\vfill \vskip 10pt \footnoterule {\footnotesize
\hoch{1} Research supported in part by DOE grant
DE-FG03-95ER40917. }

{\footnotesize \hoch{2} Research supported in part by DOE grant
DOE-FG02-84ER-40153. }


\newpage

\section{Introduction}

   There have recently been many developments in the construction of
explicit Einstein-Sasaki metrics on complete and non-singular
manifolds.  An Einstein-Sasaki space is an odd-dimensional
Einstein space that admits two Killing spinors, and thus it can be
viewed as a generalisation of the round sphere metric in the same
dimension. Einstein-Sasaki spaces are therefore important in
string theory or M-theory, since they can provide examples of supersymmetric
embeddings of AdS spacetimes. Most of the attention has
concentrated on the case of five-dimensional Einstein-Sasaki
spaces, since the generalisations of the type IIB background
AdS$_5\times S^5$ are of great importance in the AdS$_5$/CFT$_4$
correspondence \cite{maldacena}. Another important case is that of
seven-dimensional Einstein-Sasaki spaces, since these provide
supersymmetric backgrounds in M-theory that generalise
AdS$_4\times S^7$.

   For quite some time, the only explicitly-known examples of
five-dimensional Einstein-Sasaki spaces were $S^5$, which is the
homogeneous space $SO(5)/SO(4)$, and $T^{1,1}$, which is the
homogeneous space $(SU(2)\times SU(2))/U(1)$, as well as quotients
thereof.\footnote{General proofs of the existence of inhomogeneous
Einstein-Sasaki spaces were given in \cite{boygal1,boygal2}.} 
A countably infinite class of inhomogeneous examples
$Y^{p,q}$ was recently obtained in \cite{gamaspwa1}, where $p$
and $q$ are coprime positive integers. These were soon generalised
to arbitrary higher odd dimensions in \cite{gamaspwa2}, with some further
generalisations in \cite{gamaspwa3,chlupova}. A much larger class of
Einstein-Sasaki spaces in five dimensions was then constructed in
\cite{cvlupapo1}; they are denoted by $L^{p,q,r}$, where $p$, $q$
and $r$ are coprime positive integers with $0<p\le q$ and $0<r<
p+q$. The previous examples $Y^{p,q}$ arise as the special cases
$L^{p-q,p+q,p}$. Generalisations to new Einstein-Sasaki spaces
$L^{p,q,r_1,\cdots, r_{n-1}}$ spaces in $D=2n+1$ dimensions were
also given in \cite{cvlupapo1,cvlupapo2}.

According to the AdS$_5$/CFT$_4$ correspondence, five-dimensional
Einstein-Sasaki spaces are associated with four-dimensional ${\cal
N}=1$ superconformal field theories \cite{klebwitten}. These can
be described in terms of ``quiver'' gauge theories, which has been
extensively discussed, for example, in
\cite{martspar2,berbigcot,befrhamasp,marspayau,martspar,benkru}.
It has also been conjectured that the seven-dimensional
Einstein-Sasaki spaces are associated with three-dimensional
${\cal N}=2$ superconformal field theories \cite{klebwitten,fig},
although much less is known about these.

   There is a one-to-one correspondence between Einstein-Sasaki metrics
in dimension $D=2n+1$ and Einstein-K\"ahler metrics in dimension
$D=2n$ (see, for example, \cite{gibharpop} for a recent discussion of this).
Specifically, if $d\bar s^2$ is a $(2n)$-dimensional Einstein-K\"ahler
metric satisfying $\bar R_{ij} =(2n+2)\, \lambda\, \bar g_{ij}$,
then the metric
\be
ds^2 = (d\tau + A)^2 + d\bar s^2\label{ektoes}
\ee
on the $U(1)$ bundle over $d\bar s^2$ is a $(2n+1)$-dimensional
Einstein-Sasaki metric with $R_{ab}= 2n\, \lambda\, g_{ab}$, where
$dA=2\sqrt{\lambda}\, J$ and $J$ is the K\"ahler form on $d\bar
s^2$.  Thus, the local construction of $(2n+1)$-dimensional
Einstein-Sasaki metrics is equivalent to the local construction of
$(2n)$-dimensional Einstein-K\"ahler metrics.  However, a subtle
point first emphasised in the physics literature in the work of
\cite{gamaspwa1} is that the criteria for being able to extend the
local $(2n)$-dimensional Einstein-K\"ahler metric onto a complete
and non-singular manifold are much stricter than those for
extending the $(2n+1)$-dimensional Einstein-Sasaki metric onto a
complete and non-singular manifold.  To put it another way, it is
clearly the case that if $d\bar s^2$ extends onto a complete and
non-singular manifold ${\overline{\cal M}}$, then so will $ds^2$,
provided only that ${\overline{\cal M}}$ is Hodge and that the
period of $\tau$ is chosen appropriately.\footnote{A K\"ahler
manifold is Hodge if the integrals of $J$ over all 2-cycles are
rationally related, thus allowing a choice of period for $\tau$
that removes all conical singularities.}  However, it can be the
case that the Einstein-Sasaki metric $ds^2$ extends onto a
complete and non-singular manifold even though the base-space
metric $d\bar s^2$ itself has no such extension.  This feature 
played a crucial role in
the explicit construction of non-singular Einstein-Sasaki spaces
in
\cite{gamaspwa1,gamaspwa2,gamaspwa3,chlupova,cvlupapo1,cvlupapo2}.

   In this paper, we present a construction of Einstein-Sasaki
metrics in odd dimensions $D\ge 7$ that provides new examples over
and above those that have been found in
\cite{gamaspwa2,gamaspwa3,chlupova,cvlupapo1,cvlupapo2}. Our
procedure involves first constructing new classes of local
Einstein-K\"ahler metrics in all even dimensions $D\ge 6$, and then
using (\ref{ektoes}) to generate the associated local
Einstein-Sasaki metrics.  Having obtained the local metrics, we
then investigate the conditions under which they can be extended
onto complete and non-singular manifolds.  We find that, except in
rather trivial cases, the Einstein-K\"ahler metrics do not admit such
smooth extensions whereas the Einstein-Sasaki metrics do. Since a
general discussion of the circumstances under which complete and
non-singular spaces arise is quite involved, we restrict ourselves
to presenting some examples which suffice to establish the basic
principle.

   The organisation of the paper is as follows.  We begin in section 2
by presenting the local construction of the Einstein-K\"ahler
metrics in dimension $D=6$ and their lifting, {\it via}
(\ref{ektoes}), to Einstein-Sasaki metrics in $D=7$.  In section
3, we analyse the global structure of these six-dimensional and
seven-dimensional metrics.  We show that the Einstein-K\"ahler
metrics themselves generally do not extend smoothly onto complete
and non-singular manifolds, except in the special limits of either
$\CP^3$ or $\CP^2\times \CP^1$. However, we find that the
seven-dimensional Einstein-Sasaki metrics do extend onto complete
non-singular manifolds, provided that the various parameters in
the metrics are chosen appropriately. We give the general
criteria for such smooth extensions and we present some explicit
examples that establish the principle that this construction
yields a non-empty set of new examples.  In section 4, we
generalise the local construction to give new Einstein-K\"ahler
metrics in all even dimensions $D\ge 6$, and thus new
Einstein-Sasaki metrics in all odd dimensions $D\ge 7$. We do not
analyse the details of the global structures in these cases but
expect the results to be similar to those of the $D=7$
Einstein-Sasaki metrics. The paper ends with conclusions in
section 5.

\section{Local construction in $D=6$ and $D=7$}

   As discussed in the introduction, our strategy is to first
construct local expressions for Einstein-K\"ahler metrics in six
dimensions and then lift these, using (\ref{ektoes}), to give
local expressions for Einstein-Sasaki metrics in seven dimensions.
The global analysis, as well as the extension to higher
dimensions, will be given in subsequent sections.

\subsection{$D=6$ Einstein-K\"ahler metric}\label{d6eksec}

    We begin by making the following ansatz for six-dimensional
metrics:
\bea
ds_6^2 &=& \fft{(x-y) dx^2}{X} +
\fft{(x-y) dy^2}{Y} +
\fft{X}{x-y} (d\chi - \fft{y}{\beta}\,\sigma_3)^2 +
\fft{Y}{x-y} (d\chi - \fft{x}{\beta}\, \sigma_3)^2 \nn\\
&&+
\fft{xy}{\beta} (\sigma_1^2 + \sigma_2^2)\,,\label{6ans}
\eea
where $X$ is a function of $x$ and $Y$ is a function of $y$.
$\sigma_1$, $\sigma_2$ and $\sigma_3$ are the standard $SU(2)$
left-invariant 1-forms, satisfying $d\sigma_i = -\ft12 \ep_{ijk}\,
\sigma_j\wedge\sigma_k$.  We parameterise these in terms of Euler
angles $(\theta, \phi,\psi)$ in the usual way:
\be
\sigma_1 + \im\, \sigma_2 = e^{-\im\, \psi}\, (d\theta + \im\, \sin\theta
 d\phi)\,,\qquad \sigma_3 = d\psi + \cos\theta\, d\phi\,.
\ee
A straightforward calculation shows that (\ref{6ans}) is an
Einstein metric, satisfying $R_{\mu\nu}=5g^2\, g_{\mu\nu}$, if the
functions $X$ and $Y$ are taken to be
\be
X = -\fft{2\mu}{x} - \beta\,x - \alpha\, x^2 - \ft52 g^2 x^3\,,\qquad
Y = \fft{2\nu}{y} + \beta\,y + \alpha\, y^2 + \ft52 g^2 y^3\,,
\label{XY6}
\ee
where $\alpha$, $\beta$\, $\mu$ and $\nu$ are arbitrary constants.
We also find that it is K\"ahler, with the K\"ahler form given by
\be
J = dx\wedge (d\chi - \fft{y}{\beta}\, \sigma_3) +
       dy \wedge (d\chi - \fft{x}{\beta}\, \sigma_3) +
\fft{xy}{\beta}\, \sigma_1\wedge\sigma_2\,.
\ee
We can write this locally as $J=\ft12 dB$, where the 1-form $B$ is
given by
\be B =2(x + y) d\chi- \fft{2xy}{\beta}\,  \sigma_3\,. \label{A}
\ee
The K\"ahlerity of the metric is easily verified by checking that
$J_\mu{}^\nu\, J_\nu{}^\rho= -\delta_\mu^\rho$, and that $J$ is
covariantly constant.

      Although the metric is ostensibly parameterised by the four constants
$\alpha, \beta,\mu$ and $\nu$, there is a scaling symmetry under
which
\be
(x,y,\alpha,\beta,\mu,\nu)\longrightarrow (\lambda x, \lambda y,
    \lambda \alpha, \lambda^2\beta, \lambda^4 \mu, \lambda^4 \nu)\,.
\label{scalesym}
\ee
This can be used, for example, to set the parameter $\beta$ to
be either 1 or $-1$.  An alternative choice would be to use the scaling
symmetry to fix the value of $\mu$.

    For vanishing $\nu$, the metrics (\ref{6ans})
lie within a subset of the Einstein-K\"ahler metrics considered in
\cite{cvlupapo1,cvlupapo2}, which are the base metrics of the BPS
limit of the Euclideanised Kerr-de Sitter black holes found in
\cite{gilupapo1,gilupapo2}. The six-dimensional Einstein-K\"ahler
metrics and their associated seven-dimensional Einstein-Sasaki
metrics discussed in \cite{cvlupapo1,cvlupapo2} are generally of
cohomogeneity 3 and have three non-trivial parameters.  These
parameters can be taken to be $\alpha_1$, $\alpha_2$ and
$\alpha_3$, associated with the three independent rotation
parameters of the original seven-dimensional Euclideanised black
holes.  With this choice, the ``mass'' parameter $\mu$ is then
trivial and rescalable.  The overlap with the metrics we are
discussing in this section occurs if we set $\nu=0$ in
 (\ref{XY6}) and if we set $\alpha_2=\alpha_3$ in the six-dimensional
metrics in \cite{cvlupapo1,cvlupapo2}, thus reducing the
cohomogeneity from 3 to 2. It is worth emphasizing that the
generalisation away from the common overlap that we are presently
discussing, for nonvanishing $\nu$, is distinct from the
generalisation where $\alpha_1$, $\alpha_2$ and $\alpha_3$ are
unequal in \cite{cvlupapo1,cvlupapo2}, since the metrics
(\ref{6ans}) are still of cohomogeneity 2 when $\nu\ne0$. The
$\nu$ parameter can be thought of as a NUT-type parameter which
complements the mass parameter $\mu$.

The curvature invariant $R_{abcd}\,R^{abcd}$ for the
metric (\ref{6ans}) is given by
\bea R_{abcd}\,R^{abcd} &=&
  \fft{96 \mu^2(2x-y)(2x^2-2xy+y^2)}{x^6\, (x-y)^5} +
  \fft{96 \nu^2(x-2y)(x^2-2xy+ 2y^2)}{y^6\, (x-y)^5}\nn\\
&& + \fft{96(\mu y -\nu x)^2}{(x-y)^6} + 150 g^4\,. \label{6riem}
\eea
 From this, we see that there are curvature singularities when $x=y$ or
when $x$ or $y$ vanishes.

\subsection{Einstein-Sasaki metrics in $D=7$}

   Having obtained the six-dimensional Einstein-K\"ahler metrics
(\ref{6ans}), we now substitute into (\ref{ektoes}) in order to
obtain the associated seven-dimensional Einstein-Sasaki metrics.
This yields
\be ds_7^2 = (d\tau + \sqrt{\ft58} g\, A)^2 + ds_6^2\,, \ee
where $A$ is given by (\ref{A}). The metric is Einstein-Sasaki,
satisfying
\be
R_{\mu\nu}=\ft{15}4 g^2 g_{\mu\nu}\,.
\ee
Without loss of generality, we set $g^2=\ft85$ so that the
Einstein-Sasaki metric has the same Ricci tensor $R_{\mu\nu}=
   6 g_{\mu\nu}$ as that of a unit 7-sphere.  Then the metric takes the
local form
\bea
ds_7^2 &=& \Big(d\tau + 2(x+y) d\chi - \fft{2x y}{\beta} \sigma_3\Big)^2 +
             \fft{(x-y) dx^2}{X} + \fft{(x-y) dy^2}{Y}\nn\\
&&
+\fft{X}{x-y} (d\chi - \fft{y}{\beta}\,\sigma_3)^2 +
\fft{Y}{x-y} (d\chi - \fft{x}{\beta}\, \sigma_3)^2 +
\fft{xy}{\beta} (\sigma_1^2 + \sigma_2^2)\,,\label{d7es}
\eea
where
\be
X = -\fft{2\mu}{x} - \beta\,x - \alpha\, x^2 - 4 x^3\,,\qquad
Y = \fft{2\nu}{y} + \beta\,y + \alpha\, y^2 + 4 y^3\,.
\ee

   As in our discussion of the six-dimensional Einstein-K\"ahler
base in section \ref{d6eksec}, the seven-dimensional
Einstein-Sasaki metrics we have obtained here reduce, upon setting
$\nu=0$, to the subset of the metrics obtained in
\cite{cvlupapo1,cvlupapo2} corresponding to setting
$\alpha_2=\alpha_3$.  When $\nu$ is non-zero, the metrics we have
constructed here are new.

\section{Global Analysis}\label{d6globalsec}

     Our principal goal in this section is to study the global structure of
the seven-dimensional Einstein-Sasaki metrics obtained in
(\ref{d7es}) and to establish the conditions on the parameters in
order to have metrics that extend smoothly onto complete and
non-singular compact 7-manifolds.  Before doing this, we first
examine the global structure of the six-dimensional
Einstein-K\"ahler base metrics themselves and show that, except in
``trivial'' limiting cases, namely $\CP^3$ or $\CP^2\times \CP^1$,
they will necessarily have conical singularities.

\subsection{$D=6$ Einstein-K\"ahler metric}

   If a non-singular compact Einstein-K\"ahler space existed, it would 
be defined by having $x$ and $y$
run between adjacent roots $(x_1, x_2)$ and $(y_1,y_2)$ of
$X=0$ and $Y=0$ respectively.  Owing to the scaling symmetry
(\ref{scalesym}), we can set $y_1=1$ without loss of generality.
One can then parameterise the metric by the three constants $y_2$,
$x_1$ and $x_2$. The parameters $\alpha$, $\beta$, $\mu$ and $\nu$
can be expressed in terms of these roots, by using the equations
following from $X(x_i)=0$ and $Y(y_i)=0$, namely
\bea
4\mu + 2\beta\, x_1^2 + 2\alpha\, x_1^3 - 8 x_1^4=0\,,&&
4\mu + 2\beta\, x_2^2 + 2\alpha\, x_2^3 - 8 x_2^4=0\,,\nn\\
4\nu + 2\beta\, y_1^2 + 2\alpha\, y_1^3 - 8 y_1^4=0\,,&&
4\nu + 2\beta\, y_2^2 + 2\alpha\, y_2^3 - 8 y_2^4=0\,.
\eea
As can be seen from (\ref{6riem}), the metric has power-law
singularities at $x=y$, $x=0$ and $y=0$. Such singularities can be
avoided, while also having Euclidean signature, for the following
cases:
\bea \hbox{Case 1}:&& x_1>y_1,\quad x_1>y_2,\quad x_2>y_1,\quad x_2>y_2,
   \quad X>0\,,\quad Y>0\,,\quad
\beta>0\,,\nn\\
\hbox{Case 2}:&& x_1 \hbox{ and } x_2<0,\quad y_1 \hbox{ and }
y_2>0 \,,\quad
X<0\,,\quad Y<0\,,\quad \beta <0\,.\label{twocases}
\eea
With either of these choices, the coordinates $x$ and $y$ range
between endpoints in two non-overlapping intervals, thus ensuring
that none of $x-y$, $x$ or $y$ vanishes.

       While the above constraints ensure that the solution has no
power-law singularities, there can still be $\delta$-function
conical singularities at surfaces where the metric degenerates.
Specifically, these degeneracies occur at $x=x_1$ and $x_2$,
$y=y_1$ and $y_2$, and $\theta=0$ and $\pi$.  At each degeneracy,
there is an associated Killing vector $K$ whose norm $K^2 =
K^\mu\, K_\mu$ goes to zero. These are given by
\bea x=x_1: && K_1 =\fft{2x_1}{2\beta + 3\alpha\, x_1 + 16 x_1^2}
\Big( \fft{\del}{\del\chi} +
\fft{\beta}{x_1}\fft{\del}{\del\psi}\Big) \,,
\nn\\
x=x_2: && K_2 =\fft{2x_2}{2\beta + 3\alpha\, x_2 + 16 x_2^2} \Big(
\fft{\del}{\del\chi} + \fft{\beta}{x_2}\fft{\del}{\del\psi}\Big)
\,,
\nn\\
y=y_1: && K_3 =\fft{2y_1}{2\beta + 3\alpha\, y_1 + 16 y_1^2} \Big(
\fft{\del}{\del\chi} + \fft{\beta}{y_1}\fft{\del}{\del\psi}\Big)
\,,\nn\\
y=y_2: && K_4 =\fft{2y_2}{2\beta + 3\alpha\, y_2 + 16 y_2^2} \Big(
\fft{\del}{\del\chi} + \fft{\beta}{y_2}\fft{\del}{\del\psi}\Big)
\,,\nn\\
\theta=0:&&  K_5=\fft{\del}{\del\psi} - \fft{\del}{\del\phi}\,,\nn\\
\theta=\pi:&&  K_6=\fft{\del}{\del\psi}+ \fft{\del}{\del\phi}\,.
\label{d6killing}
\eea
In each case, we have normalised the Killing vector so that the
associated ``Euclidean surface gravity,'' defined by
\be
\kappa^2 = \fft{g^{\mu\nu}\, (\del_\mu K^2) (\del_\nu K^2)}{
             4 K^2}\,,
\ee
in the limit that the degenerate surface is
reached, is equal to unity.  As discussed in \cite{cvlupapo1,cvlupapo2}, 
this means
that the translation generated by the Killing vector should have
period $2\pi$ if a conical singularity is to be avoided on the
degenerate surface.

   The six Killing vectors (\ref{d6killing}) all lie in the
three-dimensional vector space spanned by $\del/\del\phi,
\del/\del\psi$ and $\del/\del\chi$.  Following the arguments given
in \cite{cvlupapo1,cvlupapo2}, this implies
that they should be linearly dependent with integer coefficients.
In particular, any three among the four Killing vectors $ K_1$,  $K_2$,
$K_3$ and $K_4$ should be linearly dependent.  For example
\be
n_1  K_1 + n_2  K_2 + n_3 K_3=0\,,
\ee
for co-prime integers $n_i$.  Such conditions can easily be satisfied,
for example, by choosing the parameters so that the roots $x_i$ and
$y_i$ are rational. However, there are further restrictions that must
be taken into account.  For example, $ K_2$ and $ K_3$ can both vanish
simultaneously, where $x=x_2$ and $y=y_1$.  This implies that the
Killing vector $K=n_2 K_2 + n_3 K_3$ also vanishes there.  It
generates translations with the period $2\pi\, \gcd(n_2,n_3)$.  Thus,
we have $n_1=\gcd (n_2,n_3)$.  Analogously, we have $n_2=\gcd (n_1,
n_3)$.  This leads to the conditions
\bea
&& K_1 \pm  K_2 = n_1^\pm  K_3\,,\qquad
 K_1 \pm  K_2 = n_2^\pm  K_4\,,\nn\\
&& K_3 \pm  K_4 = m_1^\pm  K_1\,,\qquad
 K_3 \pm  K_4 = m_2^\pm  K_2\,.
\eea
These conditions can only be satisfied in certain special
cases. Firstly, if $\mu$ and $\nu$ both vanish then the metric
becomes the standard Fubini-Study metric on $\CP^3$. Secondly,
there is a particular case, in which either $\mu$ or $\nu$ vanishes,
that corresponds to the metric on $\CP^2\times \CP^1$. The
associated seven-dimensional Einstein-Sasaki spaces are $S^7$ and
$M^{1,1,1}$, respectively. With the exception of these special
cases, the Einstein-K\"ahler metric (\ref{6ans}) is singular in
the sense that it cannot be extended onto a smooth manifold
without conical singularities.

\subsection{$D=7$ Einstein-Sasaki metric}

   We shall now demonstrate that, even though the six-dimensional
Einstein-K\"ahler base space is singular, we can nevertheless
obtain non-singular seven-dimensional Einstein-Sasaki spaces from the
local metrics (\ref{d7es}). The conditions for avoiding power-law
curvature singularities are the same as those that we discussed
previously for the base metrics. Namely, we must ensure that the
$x$ and $y$ range in non-overlapping intervals such that $x-y$,
$x$ and $y$ never vanish.  The locations of the degenerate
surfaces are also the same.  However, the Killing vectors that
vanish on these surfaces are now given by
\bea x=x_1:&&  K_1 = c_1 \Big[\fft{\del}{\del \tau} - \fft{1}{2
x_1} \Big(\fft{\del}{\del\chi} + \fft{\beta}{x_1}
\fft{\del}{\del\psi}\Big)\Big]\,,\nn\\
x=x_2:&&  K_2 = c_2 \Big[\fft{\del}{\del \tau} - \fft{1}{2 x_2}
\Big(\fft{\del}{\del\chi} + \fft{\beta}{x_2}
\fft{\del}{\del\psi}\Big)\Big]\,,\nn\\
y=y_1:&&  K_3 = \td c_1 \Big[\fft{\del}{\del \tau} - \fft{1}{2
y_1} \Big(\fft{\del}{\del\chi} + \fft{\beta}{y_1}
\fft{\del}{\del\psi}\Big)\Big]\,,\nn\\
y=y_2:&&  K_4 = \td c_2 \Big[\fft{\del}{\del \tau} - \fft{1}{2
y_2} \Big(\fft{\del}{\del\chi} + \fft{\beta}{y_2}
\fft{\del}{\del\psi}\Big)\Big]\,,\nn\\
\theta=0:&&  K_5=\fft{\del}{\del\psi} - \fft{\del}{\del\phi}\,,\nn\\
\theta=\pi:&&  K_6=\fft{\del}{\del\psi}+ \fft{\del}{\del\phi}\,,
\label{d7killing}
\eea
where
\be
c_i=\fft{4x_i^2}{2 + 3\alpha\, x_i + 16 x_i^2}\,,\qquad
\td c_i=\fft{4y_i^2}{2 + 3\alpha\, y_i + 16 y_i^2}\,.
\ee
Again, we have normalised the Killing vectors so that each has
unit Euclidean surface gravity, implying
that they must each generate translations with period $2\pi$ at the
corresponding degenerate surface.  Following the earlier discussion,
it is clear that they must satisfy
\be
b_1 K_1 + b_2 K_2 + b_3 K_3 + b_4 K_4 +
b_5( K_5 +  K_6)=0\,,\label{d7con1}
\ee
for appropriate constant coefficients $b_i$.   It is
straightforward to verify from (\ref{d7killing}) that these
constants satisfy
\be
b_1 + b_2 + b_3 + b_4 + 2b_5=0\,.
\ee
To avoid conical singularities, these constants must all be rationally related.
Without loss of generality, we can then scale them so that they are 
integers.

   We can solve (\ref{d7con1}) by considering the two conditions
\be
n_1 K_1 + n_2 K_2 + n_3 K_3 + n_4 K_4=0\,,\quad m_1 K_1 + m_2
K_2 + m_3 K_3 + m_{56} ( K_5 +  K_6)=0 \,,\label{d7con2} \ee
where $n_i$ and $m_i$ are two sets of co-prime integers. From
(\ref{d7killing}), these equations imply that
\be
n_1 + n_2 + n_3 + n_4 = 0\,,\qquad
m_1 + m_2 + m_3 + m_{56}=0\,.\label{d7con3}
\ee
Once these equations are satisfied, the constants $b_i$ can be given by
\bea
&&b_1=k_1 n_1 + k_2 m_2\,,\quad
b_2=k_1 n_2 + k_2 m_2\,,\quad
b_3=k_1 n_3 + k_2 m_3\,,\nn\\
&&b_4=k_1 n_4\,,\quad
b_5=k_2 m_{56}\,,
\eea
for arbitrary integers $k_1$ and $k_2$.  
Note that in a space of Euclidean signature, if $n$ Killing vectors
$K_i$ simultaneously vanish at a
certain degenerate surface then any linear combination of these
Killing vectors, $K=m_i K_i$, also vanishes.  In particular,
if $m_i$ are integers and $K_i$ all generate translations with 
period  $2\pi$, then
the period for $K$ is $2\pi\, \gcd(m_1\,,\cdots\,,m_n)$. In our example,
any combination of three Killing vectors from the three sets
$( K_1, K_2), (K_3,  K_4)$ and $( K_5, K_6)$ will vanish on the corresponding
surface where $x=x_1$ or $x=x_2$ and simultaneously $y=y_1$ or
$y=y_2$ and also $\theta=0$ or $\theta=\pi$. This implies
the following constraints on the integers, in order to avoid conical
singularities:
\bea
&&\gcd(k_1 n_1 + k_2 m_2\,, k_1 n_3 + k_2 m_3\,, k_2 m_{56})
=\gcd(k_1 n_2 + k_2 m_2\,,k_1 n_4\,,k_2 m_{56})\,,\nn\\
&&\gcd(k_1 n_1 + k_2 m_2\,, k_1 n_4\,, k_2 m_{56})
=\gcd(k_1 n_2 + k_2 m_2\,,k_1 n_3 + k_2 m_3\,,k_2 m_{56})\,,
\label{d7con5}
\eea
for all integers $k_1$ and $k_2$.

      If the parameters in the metric are chosen so that the conditions
stated above are satisfied, then the metric will extend smoothly onto
a complete and non-singular compact 7-manifold.

    The analysis to determine when the above regularity conditions
are satisfied is quite involved.  Rather than presenting a
complete analysis, we shall just give explicit examples which show
that non-trivial solutions do, in fact, exist. The condition
(\ref{d7con2}) implies that
\bea
\fft{(2\beta + x_1(3\alpha + 16x_1))(x_2-y_1)(y_1-y_2)}{
(x_1-x_2)(2\beta + y_1 (3\alpha + 16 y_1))(x_1 - y_2)}
&=&\fft{n_1}{n_3}\equiv r_1\,,\nn\\
-\fft{(2\beta + x_2(3\alpha + 16 x_2))(x_1-y_1)(y_1-y_2)}{
(x_1-x_2)(2\beta +  y_1 (3\alpha + 16 y_1))(x_2-y_2)}
&=& \fft{n_2}{n_3}\equiv r_2\,,\nn\\
 \fft{(2\beta + x_1 (3\alpha + 16 x_1))(x_2-y_1)y_1}{
x_1(x_1-x_2)(2\beta + y_1 (3\alpha + 16y_1))}
&=& \fft{m_1}{m_3}\equiv r_3\,,\nn\\
-\fft{(2\beta + x_2(3\alpha + 16 x_2))(x_1-y_1)y_1}{
(x_1-x_2)x_2(2\beta + y_1 (3\alpha + 16y_1))}
&=& \fft{m_2}{m_3}\equiv r_4\,,\label{d7con6}
\eea
where $r_1$, $r_2$, $r_3$ and $r_4$ are rational numbers.  These
equations place severe constraints on the existence of solutions.
Recalling that the scaling symmetry (\ref{scalesym}) allows us to set
$y_1=1$ without loss of generality, we see that the right-hand sides
of the four equations (\ref{d7con6}) have only three independent
variables.  Thus, for any given set of rational numbers $(r_1, r_2,
r_3, r_4)$ there are, in general, no solutions.  We can eliminate the
quantities $x_1$, $x_2$ and $y_2$ from (\ref{d7con6}), which gives
rise to a 26th-order polynomial $P$ in $(r_1, r_2, r_3, r_4)$
involving 1866 non-factorisable terms; we shall not present this here.

   A strategy for finding a solution is to start by selecting two
rational numbers $(r_1, r_2)$ that satisfy the conditions in
(\ref{d7con5}) for $k_1=1$ and $k_2=0$.  
Next, we substitute these into the polynomial $P$
and look for rational solutions for $(r_3, r_4)$.  If such
solutions exist, we can then check if the set $(r_1,r_2,r_3,r_4)$
satisfies the conditions in (\ref{d7con5}) for all integers $k_1$ and 
$k_2$. If it does, then we
can use (\ref{d7con6}) to determine $x_1, x_2$ and $y_2$ (since we
have set $y_1=1$).  We can then check if either of the sets of
inequalities in (\ref{twocases}) is satisfied.  If this is the
case, then we have obtained an Einstein-Sasaki metric that extends
smoothly onto a complete and non-singular compact 7-manifold.  Of course,
for many of the starting choices for $r_1$ and $r_2$ the procedure
will fail, since not all of the regularity conditions will be
satisfied. With the aid of a computer, one can repeat the
procedure for different choices of $r_1$ and $r_2$ until one finds
a solution which satisfies all of the
constraints.
\footnote{One might think that a simpler search strategy would 
be to start by choosing
rational roots $x_1$, $x_2$ and $y_2$ that satisfy either of the
sets of inequalities in (\ref{twocases}), and then obtain the
(necessarily) rational numbers $r_i$ via (\ref{d7con6}). However, the results
obtained for $(r_1,r_2,r_3,r_4)$ will typically not satisfy the conditions
(\ref{d7con5}), and in practice one finds that the search for a 
valid solution using this approach takes much longer.}

     We will explicitly present two regular solutions which we have
obtained by following the above procedure.  The first solution
satisfies the conditions for Case 1 in (\ref{twocases}), whilst
the second satisfies the conditions in Case 2. The first example
is given by
\be
x_1=\ft{13}{8}\,,\qquad
x_2=\ft{59}{24}\,,\qquad
y_1=1\,,\qquad y_2=\ft43\,,
\ee
which corresponds to
\bea
(r_1, r_2, r_3, r_4)&=&(-\ft52, -\ft12, \ft{35}{26}, \ft{81}{118})\,,\nn\\
(\alpha, \beta, \mu , \nu)&=&(-\ft{686}{39}, \ft{2326}{117},
  -\ft{45253}{18432}, -\ft{368}{117})\,.
\eea
  The six Killing vectors (\ref{d7killing})
that vanish on the degenerate surfaces
satisfy the linear relations:
\bea
5 K_1 +  K_2 - 2 K_3 - 4 K_4&=&0\,,\nn\\
2065 K_1 + 1053 K_2 + 1534 K_3 - 2326( K_5+  K_6)
&=& 0\,.
\eea
The condition (\ref{d7con5}) applied to this case becomes
\bea
\!\!\!\!\!\!&&\gcd(5k_1 + 2065k_2, -2k_1 +1534k_2, -2326k_2)=
\gcd(k_1 +1053k_2, -4k_1, -2326k_2)\,,\nn\\
\!\!\!\!\!\!&&\gcd(5k_1 + 2065k_2, -4k_1, -2326k_2)=
\gcd(k_1 +1053k_2, -2k_1 +1534k_2, -2326k_2)\,,
\eea
which is satisfied for all integers $k_1$ and $k_2$.

         The corresponding functions $X$ and $Y$ are
\bea
X&=&\fft{(8x-13)(59-24x)(2496x^2-784x-767)}{59904x}\,,\nn\\
Y&=&\fft{4(y-1)(4-3y)(92+161y-67y^2)}{117y}\,. \eea
It can be seen from Figure 1 that this example satisfies all of
the inequalities specified in Case 1 of (\ref{twocases}).

\bigskip

\begin{figure}[ht]
\epsfxsize=3.5truein \leavevmode\centering \epsfbox{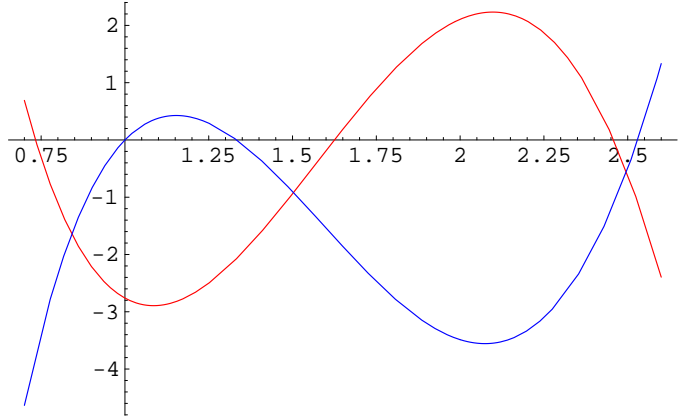}
\caption{{\it The functions $X(x)$ (in red) and $Y(y)$ (in blue),
showing the non-overlapping closed intervals in which they are
greater than or equal to zero. The horizontal axis is $x$ and $y$,
respectively, for the two functions. ($X(x)$ is heading to
$-\infty$ at large positive $x$, whilst $Y(y)$ is heading to $+\infty$ at
large positive $y$.)}}
\end{figure}

      The second example, which instead satisfies the inequalities listed in
Case 2 in (\ref{twocases}), is given by
\be
x_1=-\ft29\,,\qquad x_2=-2\,,\qquad
y_1=1\,,\qquad y_2=\ft14\,.
\ee
The parameters $\alpha$, $\beta$\, $\mu$ and $\nu$ and the
rational numbers $r_i$ are given by
\bea
(r_1,r_2,r_3,r_4)&=&(-\ft92,-\ft12,-\ft{51}{4},-\ft34)\,,\nn\\
(\alpha,\beta,\mu,\nu)&=& (\ft{35}{9},
-\ft{25}{3}, \ft29, \ft29)\,.
\eea
The six Killing vectors (\ref{d7killing}) that vanish on the
degenerate surfaces satisfy the linear relations
\bea
9 K_1 + K_2 - 2 K_3 - 8 K_4 &=&0\,,\nn\\
51 K_1 + 3K_2 - 4 K_3 - 25 (K_5 + K_6) &=&0\,.
\eea
To avoid conical singularities, it is necessary to satisfy
the condition (\ref{d7con5}), which, applied to this example,
is given by
\bea
&&\gcd(9k_1+51k_2,-2k_1-4k_2, -25k_2)=
\gcd(k_1+3k_2, -8k_1, -25k_2)\,,\nn\\
&&\gcd(9k_1+51k_2,-8k_1, -25k_2)=
\gcd(k_1+3k_2, -2k_1-4k_2, -25k_2)\,.
\eea
It is straightforward to verify that this condition is
satisfied for all integers $k_1$ and $k_2$.

     The functions $X(x)$ and $Y(y)$ in this example are given by
\bea
X&=& -\fft{2}{9x} (1-x)(1-4x) (2 + x) (2+9x)\,,\nn\\
Y&=& \fft{2}{9y} (1-y)(1-4y)(2 + y) (2 + 9y)\,. \eea
As can be seen from Figure 2, these functions are negative in the
non-overlapping ranges $x_2 \le x \le x_1$ and $y_2 \le y \le y_1$ of their
respective arguments, thus satisfying the conditions of Case 2.

\bigskip

\begin{figure}[ht]
\epsfxsize=3.5truein
\leavevmode\centering
\epsfbox{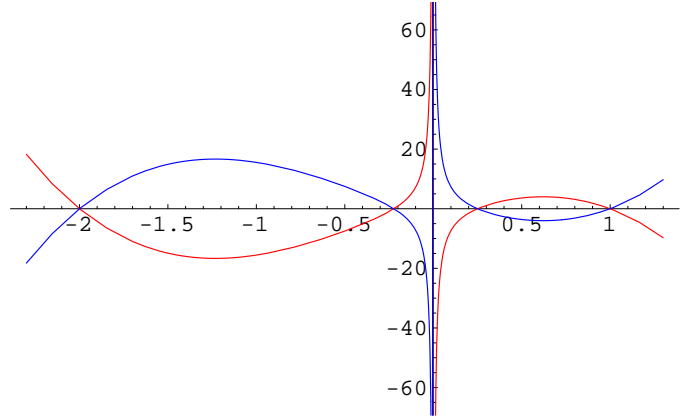}
\caption{{\it The functions $X(x)$ (in red) and $Y(y)$ (in blue),
showing the non-overlapping
closed intervals in which they are less than or equal to zero. The
horizontal axis is respectively $x$ and $y$ for the two functions.
($X(x)$ is heading to $-\infty$ at large positive $x$ and to $+\infty$ 
at large negative $x$, whilst $Y(y)$ is heading to
$+\infty$ at large positive $y$ and to $-\infty$ at large negative $y$.)}}
\end{figure}

\section{Generalisation to Arbitrary Dimension}

   We may generalise the construction of six-dimensional
Einstein-K\"ahler metrics given in section \ref{d6eksec} to arbitrary
even dimensions $D\ge 6$.  To do this, it is useful first to derive a
set of conditions for having a $(2n)$-dimensional Einstein-K\"ahler
metric $ds^2$ that is normalised, for convenience, to satisfy
$R_{ij}= 2(n+1)\, g_{ij}$. If the K\"ahler form is $J=\ft12 dB$, then
we may construct the $(2n+1)$-dimensional Einstein-Sasaki metric using
(\ref{ektoes}), and hence the $(2n+2)$-dimensional Ricci-flat K\"ahler
metric $d\tilde s^2$ on the Calabi-Yau cone over this. With the
normalisations we are using, the metric on the Calabi-Yau cone will be
given by
\be d\tilde s^2 = dr^2 + r^2 (d\tau + B)^2 + r^2 ds^2\,. \ee
It is easily verified that this has K\"ahler form
\be \widetilde J = rdr\wedge (d\tau +B) + r^2 J\,. \ee
We may also take the canonical holomorphic $(n+1)$-form
$\widetilde \Omega$ to be given by
\be
\widetilde \Omega = e^{\im\, (n+1)\tau}\, r^n\,
       [dr + \im\, r(d\tau + B)]\wedge \Omega\,,
\ee
where $n$ is the canonical holomorphic $n$-form on $ds^2$.
It can now easily be verified that the conditions $d\wtd J=0$ and
$d\wtd\Omega=0$, which ensure that $d\td s^2$ is Ricci-flat and K\"ahler,
imply that
\be dJ=0\,,\qquad d\Omega = \im\, (n+1)\, B\wedge
\Omega\,.\label{ekcon} \ee
These, then, are the conditions for the original $2n$-dimensional metric
$ds^2$ to be Einstein-K\"ahler, satisfying $R_{ij}= 2(n+1)\, g_{ij}$.

   Equipped with these equations, we now make the following ansatz for
a $(2n+4)$-dimensional Einstein-K\"ahler metric $d\hat s^2$:
\be
d\hat s^2 = \fft{x-y}{X}\, dx^2 + \fft{x-y}{Y}\, dy^2 +
 \fft{X}{x-y}\, (d\chi- y\, \sigma)^2 +
  \fft{Y}{x-y}\, (d\chi - x\, \sigma)^2 + xy\, ds^2\,,\label{hatmet}
\ee
where $X$ is a function of $x$, $Y$ is a function of $y$,
\be
\sigma=d\psi +B\,,\label{sigdef}
\ee
and $ds^2$ is an Einstein-K\"ahler metric satisfying $R_{ij}=
2(n+1)\, g_{ij}$, with K\"ahler form $J=\ft12 dB$.  If we define
\be \hat B = 2(x+y) d\chi - 2xy \, \sigma\,, \ee
then it is easily seen that
\be \hat J \equiv  \ft12 d\hat B = dx\wedge (d\chi-y\, \sigma) +
            dy\wedge (d\chi-x\, \sigma) - xy \, J\label{hatj}
\ee
defines an almost-complex structure with respect to the metric
(\ref{hatmet}).  Its manifest closure
is one of the two conditions for $d\hat s^2$ to be
Einstein-K\"ahler, with K\"ahler form given by $\hat J$.

    Next, we define the $(n+2)$-form
\be
\hat\Omega = e^{\im\, (\ft12(n+2) \alpha \chi + \gamma\psi)}\, (xy)^{n/2}\,
\epsilon_1\wedge\epsilon_2\wedge\Omega\,,\label{hatom}
\ee
where $\Omega$ is the holomorphic $n$-form on $ds^2$ (satisfying the
second equation in (\ref{ekcon})), and
\bea
\epsilon_1&=&\Big( \fft{X}{x-y}\Big)^{-1/2}\, dx - \im \,
        \Big( \fft{X}{x-y}\Big)^{1/2}\, (d\chi - y\, \sigma)\,,\nn\\
\epsilon_2&=&
\Big( \fft{Y}{x-y}\Big)^{-1/2}\, dy - \im \,
        \Big( \fft{Y}{x-y}\Big)^{1/2}\, (d\chi - x\, \sigma)\,.
\eea
It can be seen
that $\hat\Omega$ is holomorphic with respect to the almost-complex
structure defined by $\hat J$ in equation (\ref{hatj}).
The remaining condition
given in (\ref{ekcon}) for a metric to be Einstein-K\"ahler, which in
the present case becomes
\be d\hat \Omega = \im\, (n+3)\, \hat B\wedge \hat\Omega\,, \ee
is satisfied if $\gamma= 2n+2$ and the functions $X$ and $Y$ satisfy
\bea
x X' + n X + 4(n+3) x^3 + (n+2)\alpha x^2 + 4(n+1) x &=&0\,,\nn\\
y Y' + n Y - 4(n+3) y^3 - (n+2)\alpha y^2 - 4(n+1) y &=&0\,,
\eea
where a prime indicates a derivative with respect to the argument
of the functions $X(x)$ and $Y(y)$ respectively.  Thus, the metric
$d\hat s^2$ given in (\ref{hatmet}) is Einstein-K\"ahler,
satisfying $\hat R_{ab}
  = 2(n+3)\, \hat g_{ab}$,
if $ds^2$ is Einstein-K\"ahler with $R_{ij}=2 (n+1) g_{ij}$ and
the functions $X$ and $Y$ are given by
\be
X = - 4 x^3 - \alpha x^2 - 4x - \fft{\mu}{x^n}\,,\qquad
Y = 4 y^3 + \alpha y^2 + 4y + \fft{\nu}{y^n}\,.\label{XYn}
\ee
The quantities $\mu$, $\nu$ and $\alpha$ are arbitrary constants,
the K\"ahler form is given by (\ref{hatj}) and the holomorphic
$(n+2)$-form is given by (\ref{hatom}).  The Einstein-K\"ahler
metrics that we have obtained in this construction are
generalisations of the six-dimensional Einstein-K\"ahler metrics
in section \ref{d6eksec}.\footnote{Note that we could, as in
section \ref{d6eksec}, introduce an additional ``trivial''
parameter $\beta$, which would multiply the linear terms in $X$
and $Y$ in (\ref{XYn}) and divide the $x y\, ds^2$ term and the
$\sigma$ 1-forms in (\ref{hatmet}).  One can then set $\beta=\pm1$
using the scaling symmetry analogous to the one discussed in the
six-dimensional case in section \ref{d6eksec}.}  In that case, the
starting point was the canonically-normalised two-dimensional
Einstein-K\"ahler metric $\ft14 (\sigma_1^2 + \sigma_2^2)$ on
$\CP^1=S^2$, while in the generalisation (\ref{hatmet}) we instead
began with an arbitrary $(2n)$-dimensional Einstein-K\"ahler base
metric $ds^2$.

   Having obtained the $(2n+4)$-dimensional Einstein-K\"ahler metrics
(\ref{hatmet}), we can of course immediately write down the
associated $(2n+5)$-dimensional Einstein-Sasaki metrics by using
the construction given in (\ref{ektoes}).  With the normalisations
we are using here, these will be given by
\bea
d\tilde s^2 &=& [d\tau -2 (x+y) d\chi + \fft{2xy}{\beta} \sigma]^2
     + \fft{x-y}{X}\, dx^2
  + \fft{x-y}{Y}\, dy^2\nn\\
&& +
 \fft{X}{x-y}\, (d\chi- \fft{y}{\beta}\, \sigma)^2 +
  \fft{Y}{x-y}\, (d\chi - \fft{x}{\beta}\, \sigma)^2 +
\fft{xy}{\beta}\, ds^2\,,
\eea
where $\sigma$ is given by (\ref{sigdef}),
\be
X = - 4 x^3 - \alpha x^2 - 4 \beta x - \fft{\mu}{x^n}\,,\qquad
Y = 4 y^3 + \alpha y^2 + 4 \beta y + \fft{\nu}{y^n}\,.\label{XYn2}
\,,
\ee
and $ds^2$ is an Einstein-K\"ahler metric satisfying $R_{ij}=
2(n+1) g_{ij}$ and with K\"ahler form given locally by $J=\ft12
dB$. The Einstein-Sasaki metric $d\tilde s^2$ satisfies
$\widetilde R_{ab} = (2n+4) \tilde g_{ab}$. For convenience, we
have included the ``trivial'' $\beta$ parameter mentioned in
footnote 3; it can be set to $\pm1$ by using the
previously-mentioned scaling symmetry.

   It can easily be seen that if we set the parameter $\nu$ in (\ref{XYn})
to zero, the metrics reduce to special cases of those discussed
previously in \cite{cvlupapo1,cvlupapo2}, namely where all except one
of the ``rotation parameters'' $\alpha_i$ in those papers are set
equal.  The inclusion of the ``NUT'' parameter $\nu$ yields new metrics
that lie outside those previously discussed in the literature.

   It is worth remarking that the construction we have discussed here
can also be applied in the case when $n=0$, for which
(\ref{hatmet}) is a four-dimensional Einstein-K\"ahler metric,
with no $x y\, ds^2$ term. In this special case, the extra ``NUT''
parameter $\nu$ is trivial, and can be absorbed by performing a
constant shift transformation with $x\rightarrow x+ c$,
$y\rightarrow y+c$. In fact, the construction when $n=0$ merely
reproduces the metrics discussed in \cite{gamaspwa1} which, in
turn, are equivalent to the five-dimensional Einstein-Sasaki
metrics in \cite{cvlupapo1,cvlupapo2} when the ``rotation
parameters'' are set equal.  Thus, it is only for
Einstein-K\"ahler metrics in $D\ge 6$ and the associated Einstein-Sasaki
metrics in $D\ge 7$ that the new parameter $\nu$ is non-trivial.

   Using these local expressions for Einstein-Sasaki metrics, one can
again perform an analysis to find choices for the free parameters such
that the metrics will extend smoothly onto complete and non-singular
compact manifolds.  The analysis will be similar to the one we
described in detail in section \ref{d6globalsec} for the
seven-dimensional case, and we shall not discuss it further here.

\section{Conclusions}

    We have constructed new explicit
Einstein-K\"ahler metrics in all even dimensions $D=2n+4 \ge 6$,
in terms of a $(2n)$-dimensional Einstein-K\"ahler base metric.
The metrics have cohomogeneity 2 (if one chooses a homogeneous
Einstein-K\"ahler metric such as $\CP^n$ for the base) and have
the new feature of including a NUT-type parameter, along with mass
and rotation parameters. In $D\ge 8$, this construction can be
iterated to yield Einstein-K\"ahler metrics of cohomogeneity
greater than 2.

   Using a canonical construction, these metrics all yield
Einstein-Sasaki metrics of the form (\ref{ektoes}) in odd dimensions
$D=2n+5 \ge 7$. For the case $D=7$, we showed in detail that, for
suitable choices of the free parameters, the Einstein-Sasaki metrics
can extend smoothly onto complete and non-singular compact manifolds
even though the underlying Einstein-K\"ahler 6-metrics have conical
singularities. These new metrics generalise certain previously-known
countably infinite classes of Einstein-K\"ahler and Einstein-Sasaki
metrics (\ie a subset of those obtained in \cite{cvlupapo1,cvlupapo2},
which arose as supersymmetric limits of the Kerr-de Sitter metrics).
Although we have only explicitly presented two examples, it is natural
to conjecture that this construction provides a countably infinite
class of new non-singular Einstein-Sasaki spaces.

These spaces can be Wick-rotated to yield supersymmetric
Kerr-Taub-NUT-de Sitter metrics. In a forthcoming paper, it will be
shown how these solutions arise in a supersymmetric limit of more
general Kerr-Taub-NUT-de Sitter black holes \cite{chlupo}.

\section*{Acknowledgements}

   J.F.V.P. is grateful to the George P. \& Cynthia W. Mitchell Institute
for Fundamental Physics for hospitality during the course of this work.

\end{document}